\documentclass[%
 preprint,
 article,
superscriptaddress,
 amsmath,amssymb,
 aps,
]{revtex4-2}
\usepackage{lipsum}
\usepackage{graphicx}
\usepackage{dcolumn}
\usepackage{bm}
\usepackage{xcolor}
\usepackage{verbatim}

\usepackage{hyperref}

\usepackage{epstopdf}
\AppendGraphicsExtensions{.tif}
\usepackage{color,soul}
\usepackage[final]{pdfpages}
\usepackage{pgffor}

\makeatletter
\AtBeginDocument{\let\LS@rot\@undefined}
\makeatother
\begin{document}

\title{In-situ scattering studies of superconducting vacancy-ordered monoclinic TiO thin films }
\author{Merve Baksi}
 \affiliation{Department of Physics, North Carolina State University, Raleigh, NC, 27695, USA}

\author{Hawoong Hong}
\affiliation{Advanced Photon Source, Lemont, IL, 76019, USA}

\author{Divine P. Kumah}
\email{divine.kumah@duke.edu}
 \affiliation{Department of Physics, North Carolina State University, Raleigh, NC, 27695, USA}
  \affiliation{Department of Physics, Duke University, Durham NC, 27708, USA}

\date{\today}

\begin{abstract}
We investigate the structural and transport properties of vacancy-ordered monoclinic superconducting $\mathrm{TiO}$ thin films grown by molecular beam epitaxy. The evolution of the crystal structure during growth is monitored by in-situ synchrotron X-ray diffraction. Long-range ordering of Ti and O vacancies in the disordered cubic phase stabilizes the vacancy-ordered monoclinic TiO phase. The reduced structural disorder arising from vacancy-ordering is correlated with a superconductor-metal transition (SMT) in contrast to the superconductor-insulator transition (SIT) observed in cubic TiO, orthorhombic $Ti_2O_3$, and the Magneli $\gamma-Ti_3O_5$ and $\gamma-Ti_4O_7$ phase. Magnetoresistance measurements for the SIT phases indicate superconducting fluctuations persisting in the normal phase. These results confirm the role of disorder related to Ti and O vacancies and structural inhomogeneity in determining the electronic properties of the normal state of titanium oxide-based superconductors.

\end{abstract}

\maketitle

\section{Introduction}
 
Understanding the role disorder plays in modulating electronic phase transitions in quantum materials remains an active area of scientific and technological importance.\cite{giustino20212021, bouadim2011single} Manifestations of disorder in the transport properties of materials include magnetoresistive effects arising from weak and strong localization, suppressed superconductivity, thickness-dependent superconductivity, and superconductor-insulator transitions (SIT)\cite{fan2018quantum,gantmakher2010superconductor, bouadim2011single, mackenzie1998extremely, bishop1985metal, gantmakher2010superconductor}. The disorder can arise due to electronic phase separation, chemical inhomogeneity, strain, and structural phase separation driven by thermodynamic and kinetic factors. Ordered and disordered chemical and structural phase separation at multiple length scales in oxide systems have important implications on their electronic and magnetic ground states.  In the high T$_C$ superconductor YBaCuO$_3$, electronic phase separation and inhomogeneities in the distribution of oxygen vacancies lead to partially screened long-ranged interactions with the size of the domains controlling electron-phonon interactions resulting in a phonon renormalization \cite{kremer1992percolative, letz1994percolative}. 

A model system for investigating disorder-induced phase transitions is the binary titanium oxide ($Ti_xO_y$) system \cite{reed1972superconductivity, doyle1968vacancies, li2021single, feng2020phase}. $Ti_xO_y$ can crystallize in a wide range of crystal structures that depend on the Ti/O ratio and growth conditions \cite{kim2014stable, andersson2005thermodynamics}. $Ti_xO_y$ bulk and thin films exhibit a wide range of electronic properties ranging from metallic and insulating phases to superconductivity \cite{reed1972superconductivity, li2018observation, ozbek2022superconducting}. The electronic properties are directly linked to the crystal phase and stoichiometry. Recent reports of superconductivity in thin $Ti_xO_y$ films have shown the occurrence of superconducting-insulator transitions (SIT) with the onset of superconductivity increasing up to 8 K with increasing film thickness \cite{fan2018quantum, li2018orthorhombic, ozbek2022superconducting, kurokawa2017effects, zhang2017enhanced, yoshimatsu2017superconductivity, li2021single}. The SIT and the thickness-dependent superconducting transition are indicative of disorder in modulating the electronic properties of the system \cite{gantmakher2010superconductor, sacepe2008disorder, harris2018superconductivity}. In s-wave superconducting systems, as is the case for NbN thin films, an increase in disorder from the weak regime to the strong limit with decreasing film thickness, is accompanied by both a suppression in Tc and the emergence of an SIT.\cite{chand2012phase}. A similar phenomenon has been reported in Ti$_x$O$_y$ films where disorder may arise from the coexistence of polymorphs and the inhomogeneous distribution of Ti and/or O vacancies.\cite{ozbek2022superconducting, kurokawa2017effects, hulm1972superconductivity, mclachlan1982new, graciani2005role, fan2018quantum}

\begin{figure*}[ht]
\centering
\includegraphics[width=0.85\textwidth]{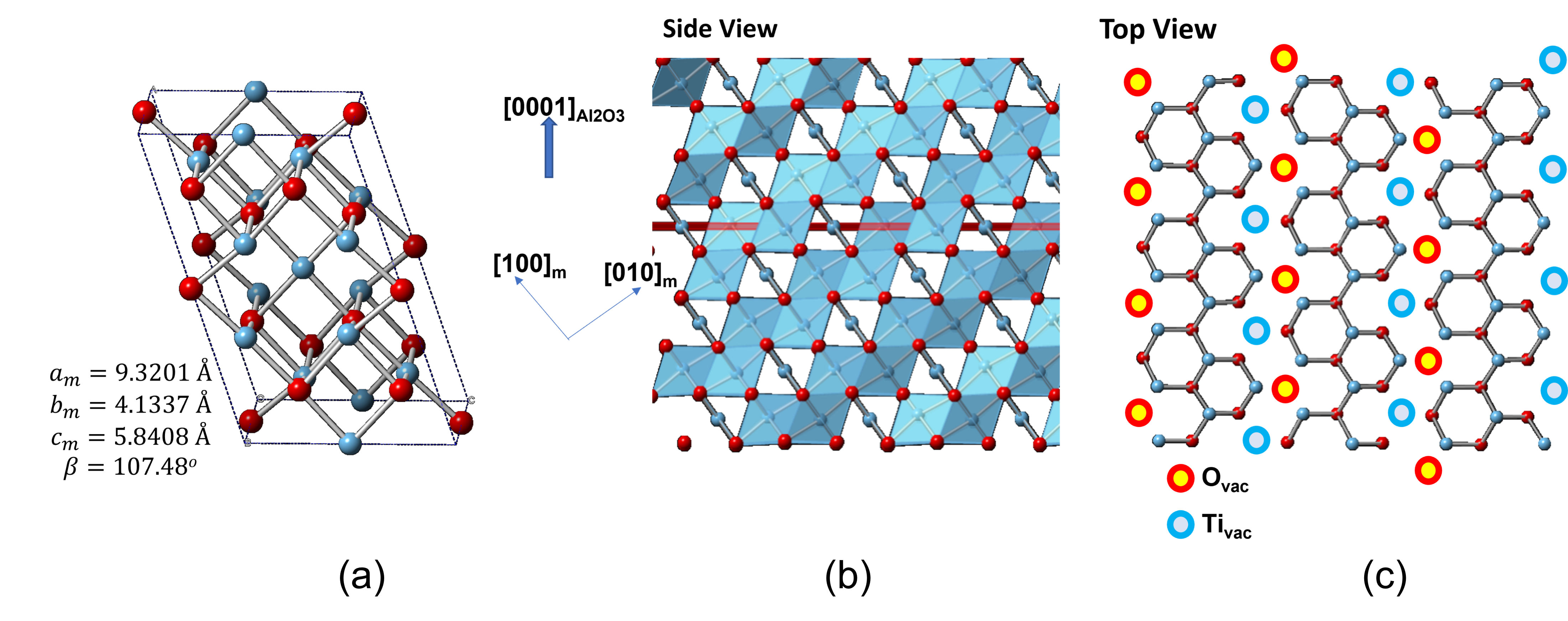}
\caption{(a) Unit cell of bulk monoclinic m-TiO. (b) Atomic structure of monoclinic m-TiO along projections in the (left) [10$\bar{1}$0]  and (right) [0001] $\alpha-Al_2O_3$ directions. (c) Ordered Ti and O vacancies are present in alternating Ti and O layers along the [0001] $\alpha-Al_2O_3$.}
\label{fig:schematic}
\end{figure*}

Identifying and suppressing the sources of disorder is critical for understanding and enhancing the superconducting properties of $Ti_xO_y$ thin films. The atomic-scale compositional control afforded by the molecular beam epitaxy (MBE) technique allows for controlling disorder in thin films. For MBE-grown $Ti_xO_y$ films, the growth temperature and the ratio between the Ti flux and the oxygen partial pressure determine the composition, structural phase, and transport properties of $Ti_xO_y$ films grown on (001)-oriented Al$_2$O$_3$ \cite{ozbek2022superconducting, walker1992high, alexander1990tiox, li2021single}. In the cubic TiO phase, the oxygen stoichiometry can vary from 0.85 to 1.25 with a disordered distribution of oxygen vacancies \cite{doyle1968vacancies}. When both oxygen vacancies ($O_{vac}$) and Ti vacancies ($Ti_{vac}$) are present in the disordered cubic phase at high temperatures, a transition to the vacancy-ordered monoclinic phase at the ordering temperature of 1026 $^o$C for $Ti_{vac}/O_{vac}=1$ is observed. The monoclinic (\textit{B2/m}) structure is characterized by periodic Ti and O vacancies in every third plane (100)$_m$ plane in the [010]$_m$ direction \cite{terauchi1978vacancy, reed1972superconductivity, watanabe1966ordered, watanabe1967ordered} (The \textit{m} subscript denotes the monoclinic lattice). In the monoclinic phase, there are three in-equivalent $Ti^{2+}$ and $O^{2-}$ sites. Two of the $Ti^{2+}$ form TiO$_5$ edge and corner-sharing octahedra. The third $Ti^{2+}$ is bonded to four $O^{2-}$ in a square co-planar geometry. A schematic of the monoclinic TiO unit cell is shown in Figure \ref{fig:schematic}. When grown on (0001)-oriented Al$_2$O$_3$, the (310)$_m$ planes are perpendicular to the film-substrate interface.

In this work, we utilize \textit{in-situ} X-ray diffraction measurements\cite{yan2020situ} and \textit{ex-situ} transport measurements to investigate the impact of vacancy ordering in monoclinic TiO (m-TiO) on the carrier localization and superconductivity in thin $Ti_xO_y$ films grown by MBE. \textit{In-situ} synthesis and synchrotron diffraction measurements are employed to examine the atomic-scale structural changes of the film while synthesizing the vacancy-ordered m-TiO phase. The m-TiO films display a superconducting-metal transition (SMT), which suggests the critical role of vacancy ordering in the modulating superconductivity in Ti-based superconductors.


\section{Results and Discussion}
\subsection{Film Growth}
$Ti_xO_y$ thin films were synthesized under conditions optimized for m-TiO formation in the \textit{in-situ} MBE system at the 33ID-E beamline at the Advanced Photon Source \cite{yan2020situ, hong2022situ}. The Ti flux was measured by a quartz crystal monitor and the oxygen partial pressure was determined by an ion gauge in the growth chamber. The substrate temperature, measured with an optical pyrometer, was fixed at 700 $^o$C during the film growth with an O$_2$ partial pressure of 1.2$\times 10^{-8}$ Torr. Higher oxygen pressures led to the formation of trigonal $Ti_2O_3$.\cite{ozbek2022superconducting} The $Ti_xO_y$ films were grown on 10 mm x 10 mm (001)-oriented single crystal Al$_2$O$_3$ substrates at a growth rate of 1 TiO monolayer (ML) (2.4 \AA{}/minute). The surface crystallinity was monitored during the growth by \textit{in-situ} reflection high energy electron diffraction (RHEED). The \textit{in-situ} grown films were capped after growth with a protective amorphous Al$_2$O$_3$ layer. Representative RHEED images before and after the capping layer are shown in Figure S1 of the Supplementary Materials \cite{Supplement}. During the growth, crystal truncation rods (CTRs) \cite{disa2020high} were measured around the (0006) Al$_2$O$_3$-substrate Bragg peak to determine the evolution of the film's out-of-plane lattice spacing with film thickness. Off-specular reciprocal space maps were measured after growth to confirm the m-TiO structure. The diffraction intensities were measured with a 2D Pilatus 100 K detector at a fixed incidence X-ray energy of 15.5 keV. 

\subsection{Nucleation and Relaxation}
\begin{figure}[h]
\centering
\includegraphics[width=0.8455\textwidth]{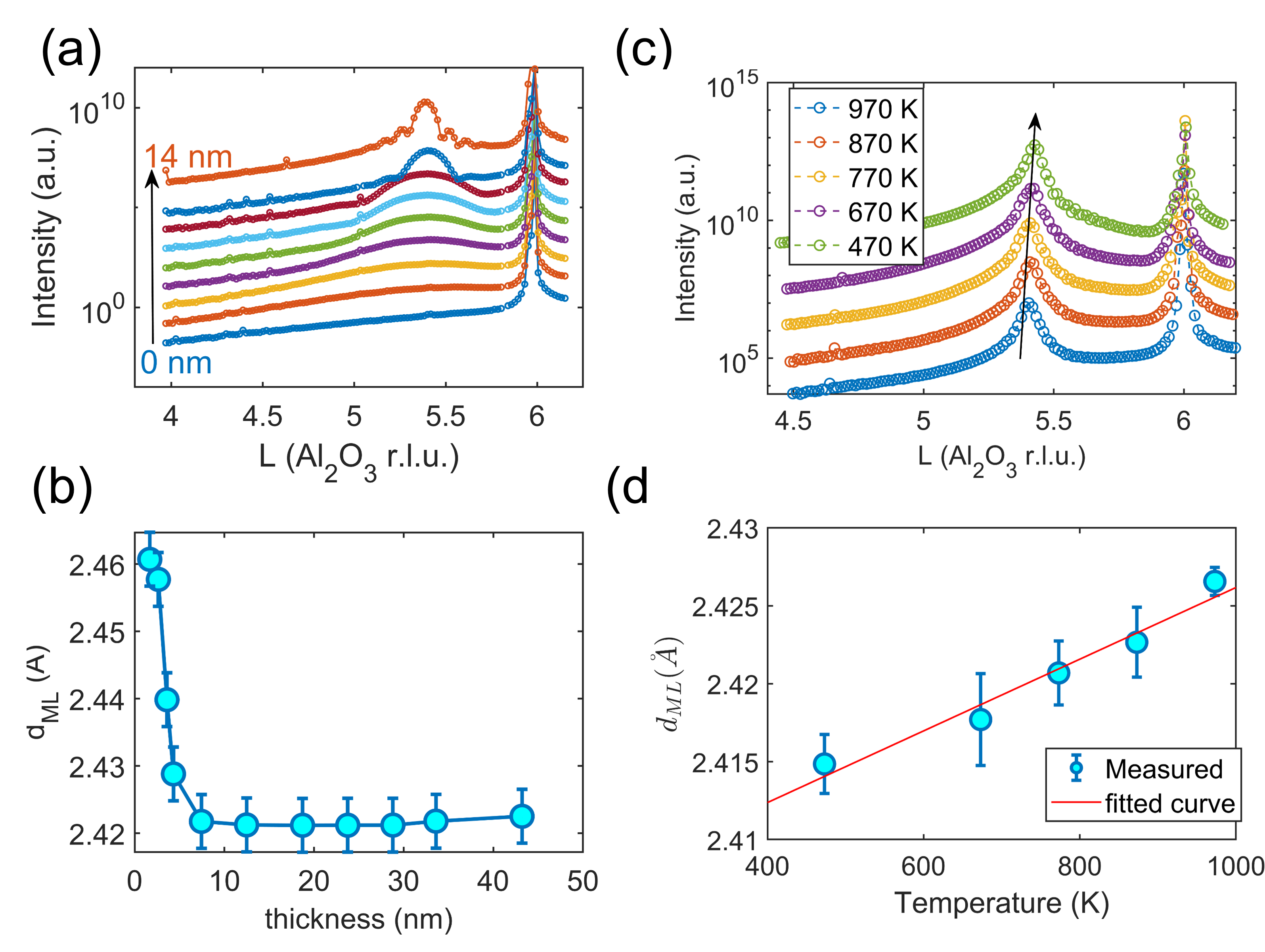}
\caption{(a) Evolution of the (00L) crystal truncation rod (CTR) during the initial growth stages of 14 nm thick m-TiO by molecular beam epitaxy. (b) The TiO monolayer spacing as a function of film thickness is determined from the diffraction measurements in (a). (c) Evolution of the (00L) CTR as the film is cooled from the growth temperature. The arrow indicates the shift in the position of the monoclinic (310)$_m$ peak of the film. (d) Out-of-plane m-TiO monolayer spacing as a function of temperature. The solid line represents a linear fit to determine the coefficient of linear expansion.}
\label{fig:lattice_constant}
\end{figure}

The evolution of the specular 00L CTR during the growth of the first 14 nm of the m-TiO film is shown in Figure \ref{fig:lattice_constant}(a). The film Bragg peak intensity increases and finite thickness oscillations develop as the film thickness increases. The period of the finite thickness oscillations is consistent with the film growth rate of 1 TiO ML/min. The film Bragg peak is related to the $Ti_xO_y$ plane spacing, $d_{ML}$, along the growth direction, where $d_{ML}=c_{Al_2O_3}/L$. $c_{Al_2O_3}$ is the lattice constant of the substrate at the growth temperature and \textit{L} is in reciprocal lattice units (r.l.u.) of the Al$_2$O$_3$ substrate. Figure \ref{fig:lattice_constant}(b) shows the changes in the monolayer spacing with film thickness. $d_{ML}$ evolves from 2.46 $\AA{}$ to 2.42 $\AA{}$ after 8 nm and remains constant till the growth is terminated after 44 nm. 

The changes in the lattice spacing can arise for multiple reasons. First, coherent strain relaxation can lead to a gradual change in the lattice parameter. Due to the lattice mismatch between (310)$_m$-oriented m-TiO (3d$(_{3\bar{2}0})_m$=5.08 \AA{} at 300 K) and Al$_2$O$_3$ (a$=4.78$ \AA{}), the film will be under $5.3\%$ compressive strain. The RHEED spacings relax from the substrate spacing to the film spacing after ~2 ML (0.4 nm),\cite{ozbek2022superconducting} thus, the decrease in the out-of-plane lattice parameter over a length scale of 8 nm may not be entirely due to strain relaxation. Secondly, the decrease in the $d_{ML}$ may be related to the O/Ti ratio in the cubic phase \cite{hulm1972superconductivity} where the lattice parameter contracts as the O/Ti ratio increases from 0.7 (4.198 \AA) to 1.3 (4.168 \AA{}). Assuming an initial layer of the c-TiO phase, and accounting for the thermal expansion at the growth temperature, the observed lattice constant variation is consistent with an increase in the oxygen content (decrease in Ti fraction) as the film thickness increases until an equilibrium 1:1 Ti:O value is reached after approximately 8 nm of growth.

\subsection{Structural Changes during Cooling}
As the film is cooled from $T_{growth}$ to room temperature in vacuum, the film and substrate Bragg peaks shift to higher L values as the lattice contracts as shown in Figure \ref{fig:lattice_constant}(c). Figure  \ref{fig:lattice_constant}(d) shows the changes in $d_{ML}$ for the 44 nm film as a function of temperature. The coefficient of thermal expansion, $\alpha$ is determined from the slope of the plot to be $8.3\pm1\times 10^{-6}$ $/K$. The measured value is similar to previous reports for bulk c-TiO($\alpha=6.6\times 10^{-6}$ $/K$) \cite{taylor1971thermal}.

\begin{figure}[h]
\centering
\includegraphics[width=0.85\textwidth]{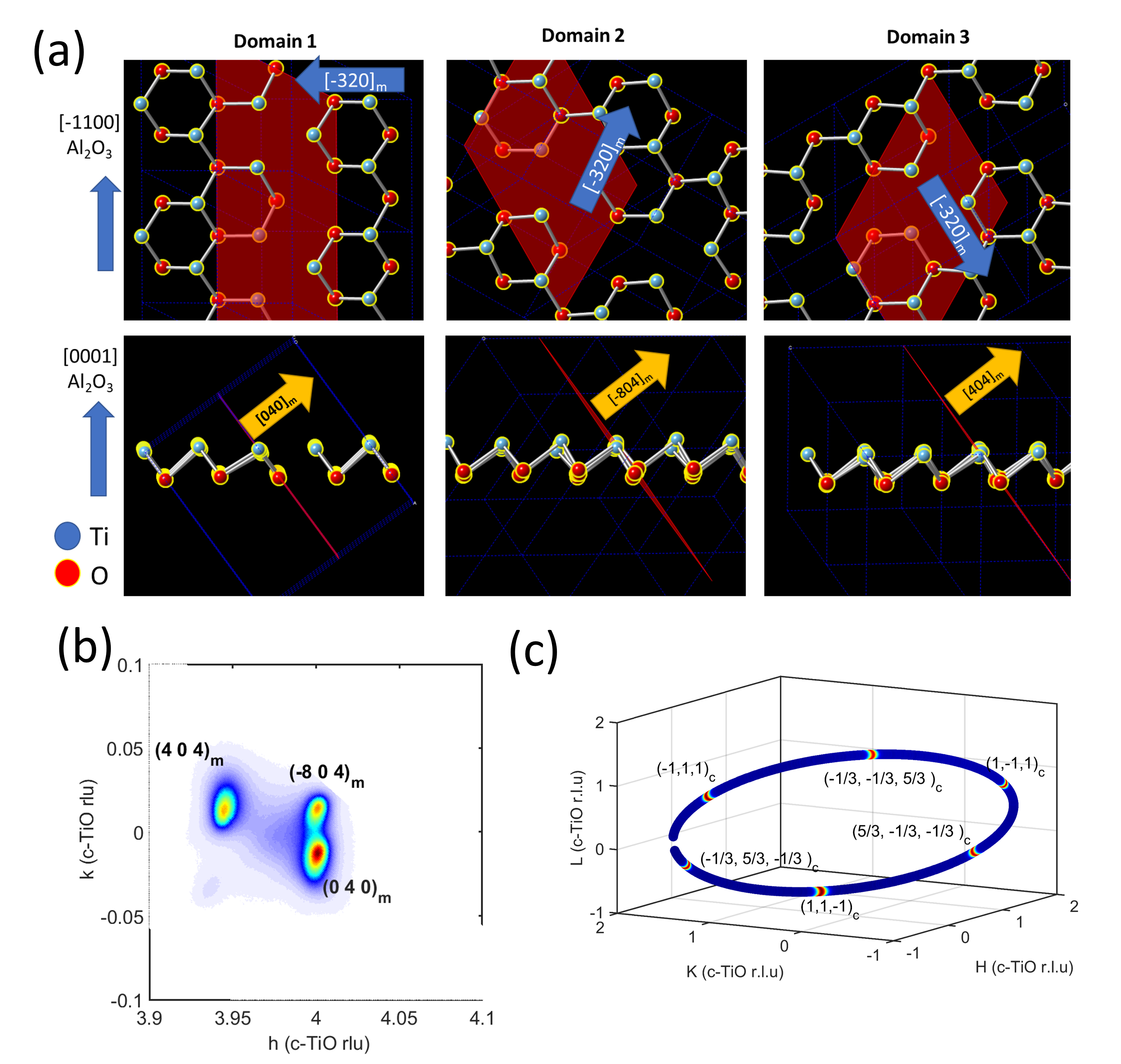}
\caption{(a) m-TiO domains on Al$_2$O$_3$. Projection along the growth direction (top panel) and corresponding Side-view (lower panel) showing the (0 4 0)$_m$, (-8 0 4)$_m$, and (4 0 4)$_m$ planes (red outlines). The domains are related by a rotation of 120$^o$ about the sample surface normal. (b)Reciprocal space map around (040)$_c$ Bragg peak for a TiO film on (0001)-oriented Al$_2$O$_3$. The splitting of the peak is due to the monoclinic distortion with $\beta=107.57^o$. (c)Three-dimensional plot of azimuthal scan around the nominal c-TiO (-1,1,1)$_c$ Bragg peak.   }
\label{fig:domain}
\end{figure}

To confirm the monoclinic symmetry of the films, reciprocal space maps are measured around the (040)$_c$ Bragg peak at room temperature (The \textit{c} subscript refers to the cubic c-TiO lattice). The monoclinic distortion will result in a splitting of the cubic (040)$_c$ Bragg peak since domains with a rotation of 120$^o$ about the surface normal are in-equivalent in the monoclinic phase. The domains are illustrated in Figure \ref{fig:domain}(a).
Figure \ref{fig:domain}(b) shows a reciprocal space map around the (040)$_c$ Bragg peak showing the monoclinic splitting. The calculated lattice parameters for the film are $a_m=9.22 \AA{}$, $b_m= 4.14\AA{}$, $c_m= 5.99 \AA{}$ and $\beta= 107.57^o$ in good agreement with bulk m-TiO \cite{reed1972superconductivity}.
Due to the periodicity of the Ti and O vacancies (Figure \ref{fig:schematic}(c)), additional peaks are measured at (\textit{m/3, n/3, o/3})$_c$ where \textit{m,n} and \textit{o} are integers as shown in Figure \ref{fig:domain}(c).

\subsection{Transport properties}
\begin{figure*}[th]
\centering
\includegraphics[width=0.95\textwidth]{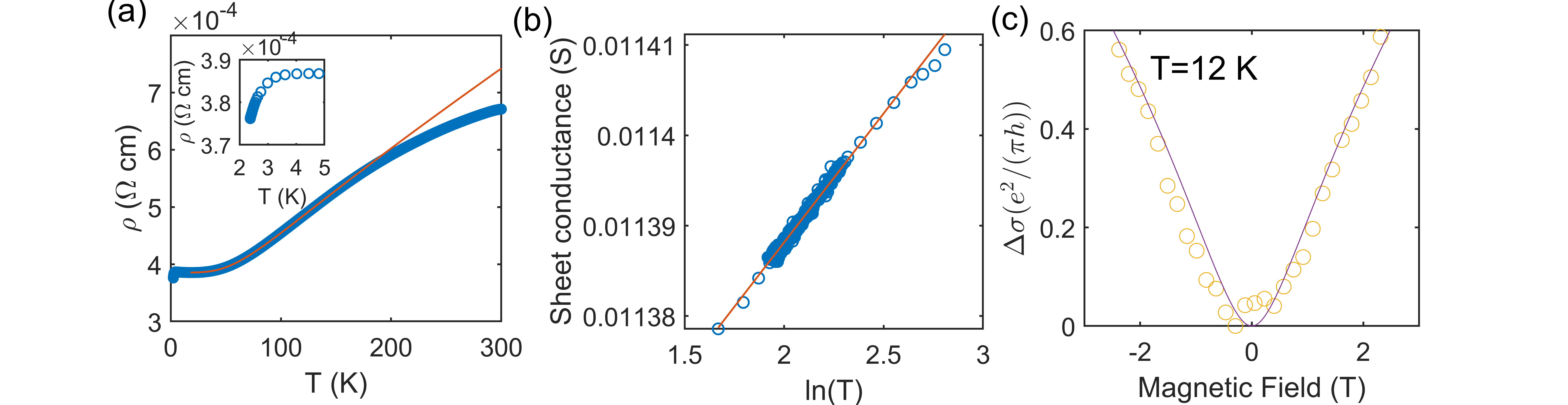}
\caption{(a) Resistivity as a function of temperature for a 44 nm m-TiO film. The inset shows the onset of superconductivity at 2.8 K. The solid red line shows the fit to the resistivity. (b) Sheet conductance versus temperature (open symbols) in the region of weak localization. The slope of the linear fit (solid line) is $pe^2/(\pi h$) where p=2.3. (c) Measured magnetoconductance (circles) at 12 K with the magnetic field applied parallel to the film surface normal. The solid line is the fits using Equation S1 of the supplemental materials with $l_{in}=20 nm $ \cite{Supplement}  }
\label{fig:transport}
\end{figure*}

The transport properties of the m-TiO film are measured \textit{ex-situ} between 300 K and 2 K in the 4-point van der Pauw configuration with Au contacts sputtered to the corners of the 10 mm x 10 mm substrate. The resistivity as a function of temperature is shown in Figure \ref{fig:transport}(a). The film is metallic with $d\rho/dt>0$ between 300 K and 18 K. A slight uptick in the resistivity is observed below 18 K indicative of weak disorder \cite{lee1985disordered}. The onset of superconductivity,$T_C^{onset}$, defined as the temperature at which the resistivity deviates from the normal-state behavior occurs at $\sim 2.8 $ K. We confirm that the transition is related to superconductivity by magnetic field-dependent resistivity measurements for a 5mm x 5mm sample grown under identical conditions (Supplementary Materials Figure S2)\cite{Supplement}.  

The normal-state resistivity $\rho$, between the onset of the weakly localized regime (18 K) and 140 K, where a change in slope of the resistivity occurs, is fitted using the Bloch-Grüneisen (BG) function. The BG function takes into account the contribution of electron scattering from acoustic phonons to the resistivity \cite{lin2004low}.
\begin{equation}
\rho = \rho_0 + A (\frac{T}{\theta_D})^n \int_0^{\theta_D/T}\frac{x^n}{(e^x-1)(1-e^{-x})}dx
\label{eqn:rho}
\end{equation}
\\
where $\theta_D$ is the Debye temperature, $\rho_0 $ is the residual resistivity and $A$ is a temperature independent constant. The exponent $n$ is related to the scattering mechanism where $n=5$ for scattering due to acoustic phonons and $n=3$ for scattering due to optical phonons related to the s-d bonding of the oxygen ions and the transition metal. Figure \ref{fig:transport}(a) shows a fit using Equation \ref{eqn:rho}. The fitted values of $n$, and $\theta_D$ are 5.08$\pm$0.35 and 350$\pm$ 5 K, respectively indicating the dominant role of scattering from low-energy acoustic phonons. The Debye temperature is comparable to the theoretical prediction of $\theta_D= 335.5$ K for m-TiO with a predicted $Tc$ of 2.8 K\cite{hosseini2021electron}.

The sheet conductance, $\sigma$, in the weakly localized regime (3 K $< T <$ 18 K) has a ln(T) dependence as shown in Figure \ref{fig:transport}(b).  For disordered 2D systems, the conductance is given by
\begin{equation}
\sigma = pe^2/(\pi h)ln(T) +c
\label{eqn:sigma}
\end{equation}

Here, $e$ is the electron charge, and \textit{p} is a coefficient related to the inelastic mean free path $l_{in}$. The inelastic mean free path scales as $T^{-p/2}$. From the slope of the $\sigma$ vs $ln(T)$ in Figure \ref{fig:transport}(b), \textit{p} is determined to be 2.3$\pm0.1$. The fitted value is close to 2 which is indicative of weak disorder in a 2-dimensional system arising from electron-electron interactions \cite{abrahams1981quasiparticle, lee1985disordered, kumah2014tuning}. Weak localization is confirmed by a negative magnetoresistance (positive magnetoconductance) as shown in Figure \ref{fig:transport}(c). The ln(T) dependence of conductivity is also expected for granular systems, however, the good agreement between the measured data and the model for 2D localization suggests that the inelastic mean free path may be comparable to the film thickness. The mean free path, determined by fitting the magnetoresistance at 12 K (Figure \ref{fig:transport}(c) is 20 nm. Thus, the system is close to a 2D-3D dimensionality crossover.  

To confirm that the system is in the weakly disordered regime, we calculate the Mott-Ioffe-Regel parameter, $k_Fl$, from the relation
 \begin{equation}  
 k_Fl=(3\pi^2)^{2/3}\hbar R_H(300 K)^{1/3}/(\rho(300 K) e^{5/3})
\label{eqn:kfl}
\end{equation}

Here, $R_H$ is the Hall resistivity, and $\rho$ is the normal state resistivity at 300 K. The values at 300 K are chosen to minimize the contribution of electron-electron interactions to the mean free path.\cite{chand2012phase} From Hall measurements on the sample ($n=2.4\times 10^{28} m^{-3}$, $\mu = 1.36cm^2/(Vs)$)),  $R_H=2.57\times 10^{-4} cm^3/C$ and $k_Fl$=7.8, indicative of low disorder ($2<k_Fl<10$) in the film. The measured Hall parameters at 300 K and 10 K are summarized in Table S1 of the Supplementary Materials \cite{Supplement}. 

A higher $T_{c, onset}$ would be expected for the vacancy-ordered m-TiO film compared to other more disordered $Ti_xO_y$ phases. However, considering that superconductivity is related to the electron-phonon coupling and the crystal structure, the $T_c$ is determined by the electron-phonon coupling constant. Theoretical predictions for $\lambda_{eff}$ ($T_{c}s$) for stoichiometric m-TiO, c-TiO, and $\gamma-Ti_3O_5$ are 0.57 (2.8 K) 0.78 (7.4 K) and 0.75 (8 K), respectively and are consistent with the lower Tc we observed for the m-TiO phase \cite{hosseini2021electron}. The enhanced Tc of 11 K in c-TiO/TiO$_{1+\delta}$ core-shell nanoparticles is attributed to interface effects \cite{xu2018nano}, thus, strain and doping can potentially induce enhanced Tcs in m-TiO thin films.\cite{sun2021first}

\begin{figure*}[th]
\centering
\includegraphics[width=0.95\textwidth]{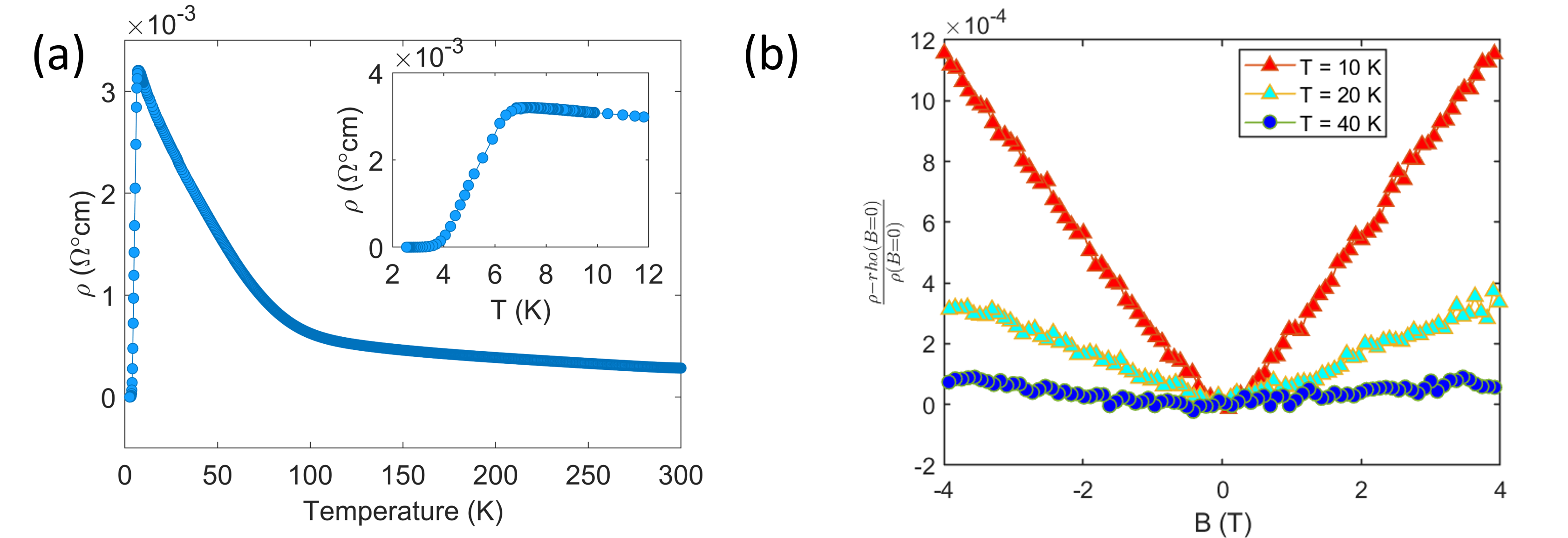}
\caption{(a) Resistivity as a function of temperature for a 40 nm c-TiO film. The inset shows the low-temperature resistivity (b) Temperature-dependent magnetoresistance with the magnetic field applied parallel to the film surface normal. }
\label{fig:cubicTransport}
\end{figure*}

In contrast to the m-TiO films which exhibit a negative MR in the normal state, the MR for cubic c-TiO film with an SIT is positive. Figure \ref{fig:cubicTransport} (a) shows the resistivity as a function of temperature for a 40 nm c-TiO film with Tc onset at $\sim7$ K. The kink in the resistivity at 100 K and the higher Tc are consistent with previous reports for PLD-grown TiO films.\cite{zhang2017enhanced} For the c-TiO film, $k_Fl=1.1$ ($n=9\times 10^{28} m^{-3}$, $\mu = 0.38 cm^2/(Vs)$), indicative of increased disorder relative to the m-TiO film. The MR is positive up to 40 K as shown in Figure \ref{fig:cubicTransport} (b) for the c-TiO film. A positive MR in the normal state for disordered superconductors has been attributed to superconducting fluctuations which survive up to a characteristic temperature T* where the superconducting gap vanishes.\cite{harris2018superconductivity, chand2012phase}  Further studies are needed to determine if superconducting fluctuations and a superconducting gap persist in the normal state of TiO-based superconductors up to T*.

\section{Conclusion}

In conclusion, we observe an SMT in vacancy-ordered m-TiO thin films grown by molecular beam epitaxy. The SMT observed is in contrast to the SIT observed in c-TiO, o-Ti$_2$O$_3$, and superconducting Magneli phases where structural and chemical disorder dominates the transport properties of the system in the normal state. The ordering of Ti and O vacancies in the monoclinic phase leads to an increase in the electronic mean-free path. Based on the current results, we postulate that approaches to dynamically modulate disorder, for example, via ionic liquid gating \cite{tian2015situ}, can potentially drive electronic transitions in the system. Additionally, superconducting fluctuations are known to survive in the normal state close to critical disorder \cite{chand2012phase, sacepe2008disorder, sacepe2011localization, fan2018quantum, dubi2007nature, chen2018carrier}, hence, understanding and controlling disorder in the m-TiO system will lead to the observation of novel electronic phases with potential applications in quantum computing and energy-efficient electronic devices.

\section*{Acknowledgments}
The authors acknowledge financial support by the US National Science Foundation under Grant No. NSF DMR-1751455. Use of the Advanced Photon Source was supported by the U.S. Department of Energy, Office of Science, Office of Basic Energy Sciences, under Contract No. DE-AC02-06CH11357. 

\section{Data Availability Statement}
The data that support the findings of this study are available from the corresponding author upon reasonable request.


\newpage
\setcounter{page}{1}

\newpage


\section*{Supplementary material}

# STRUCTURAL CHARACTERIZATION OF MONOCLINIC TITANIUM OXIDE

Bulk monoclinic (B2/m space group) TiO has lattice parameters $a_m = 9.32 Å \approx \sqrt{5} a_c$, $b_m = 4.14 Å$, $c_m = 5.85 Å \approx \sqrt{2} a_c$ and $\alpha = 107.48^o$ where the cubic lattice constant, $a_c$=4.14 Å.[1]

The transformation from the cubic real-space lattice vectors $\{\vec{a}_c\}$ to the monoclinic lattice vectors $\{\vec{a}_m\}$ is given by

$$\vec{a}_m = \vec{a}_c + 2 * \vec{b}_c$$

$$\vec{b}_m = \vec{a}_c$$

$$\vec{c}_m = \vec{a}_c - \vec{a}_c$$

In reciprocal space, $\vec{q}_c = h_c \vec{a}_c^* + k_c \vec{b}_c^* + l_c \vec{c}_c^*$. The monoclinic reciprocal lattice units are

$$\vec{a}_m^* = \frac{\vec{b}_m \times \vec{c}_m}{\vec{a}_m \cdot (\vec{b}_m \times \vec{c}_m)} = 1/3(\vec{a}_c^* + \vec{k}_c^*)$$

$$\vec{b}_m^* = \frac{\vec{c}_m \times \vec{a}_m}{\vec{a}_m \cdot (\vec{b}_m \times \vec{c}_m)} = 1(\vec{c}_c^*)$$

$$\vec{c}_m^* = \frac{\vec{a}_m \times \vec{b}_m}{\vec{a}_m \cdot (\vec{b}_m \times \vec{c}_m)} = 1/3(2\vec{h}_c^* - \vec{k}_c^*)$$

Hence, the transformation of the Miller indices $(hkl)_m --> (hkl)_c$ is given by

$$h_m = h_c + 2k_c$$

$$k_m = l_c$$

$$l_m = h_c - k_c$$

The films are grown on (0001)-oriented $Al_2O_3$ substrates. The $(310)_m$ m-TiO planes are parallel to $(0001)_{Al_2O_3}$. Due to the ordered Ti and O vacancies, three rotational domains are expected for the m-TiO film as shown in Figure 3(a) with $[\bar{3}20]_m // [\bar{1}\bar{1}20]_{Al_2O_3}$ (domain 1), $[\bar{3}20]_m // [10\bar{1}0]_{Al_2O_3}$ (domain 2), and $[\bar{3}20]_m // [01\bar{1}0]_{Al_2O_3}$ (domain 3). The domains are related by a rotation of $120^o$ about the sample surface normal.

In the cubic phase, reflections of type $(m/3, n/3, o/3)$, where m,n, and o are integers, are forbidden. These reflections are however present in the monoclinic phase. Figure 3(c)

shows an azimuthal scan about the (-1 1 1)$_c$ reflection. In addition to the three fold -1 1 1$_c$ reflections, Bragg peaks are observed at (-1/3, -1/3, 5/3)$_c$, (5/3, -1/3, -1/3)$_c$, and (-1/3, 5/3, -1/3)$_c$ corresponding to the monoclinic (-1, 5/3, -2)$_m$, (1,-1/3,2)$_m$, (3,-1/3, -2)$_m$ Bragg peaks, respectively.

**TRANSPORT MEASUREMENTS**

Hall measurements were performed in a Van der Pauw geometry at 300 K and 10 K. The results are summarized in Table I. The superconducting transition is suppressed by applying a magnetic field perpendicular to the sample surface as shown in Figure S2.

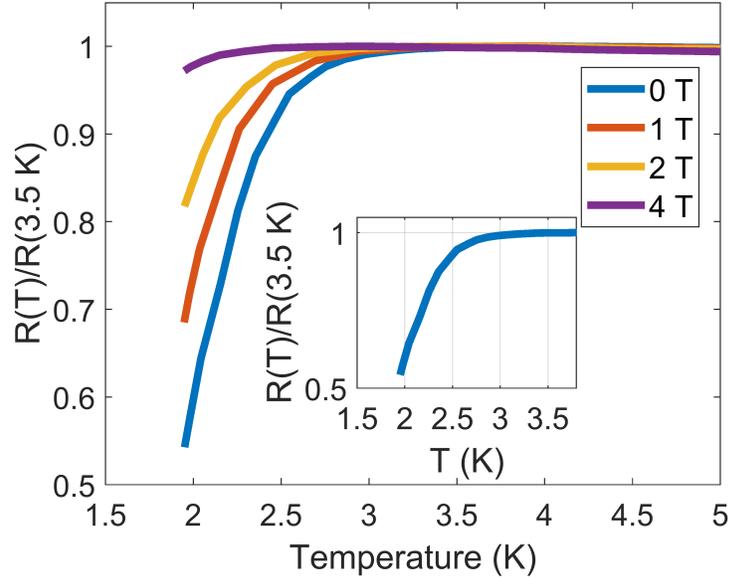

**Figure S 2.** Field-dependence of resistance as a function of temperature for a 45 nm m-TiO film on (0001)-oriented $Al_2O_3$.

**Magnetoconductance fit**

The corrections $\Delta\sigma(H)$ to the conductance for applied perpendicular magnetic fields, $H$, in disordered two-dimensional systems are given by [2]

$$\Delta\sigma(H) = \sigma(H) - \sigma(H=0) = \frac{e^2}{\pi h}[\Psi(0.5 + x^{-1}) + ln(x)] \tag{1}$$

3

| Temperature | 300 K | 10 K |
|---|---|---|
| $R_{Hall}$ [$cm^3/C$] | $2.57 \times 10^{-4}$ | $7.83 \times 10^{-4}$ |
| $n$ [$1/cm^3$] | $2.43 \times 10^{22}$ | $7.97 \times 10^{21}$ |
| $\mu$ [$cm^2/V.s$] | 1.36 | 7.08 |

**TABLE S I.** Hall coefficient, $R_{Hall}$, carrier concentration, $n$ and mobility, $\mu$ at 300 K and 10 K for the 44 nm m-TiO sample.

where $\Psi$ is the Digamma function and $x = \frac{4e}{\hbar}l_{in}^2 H$ and $l_{in}$ is the inelastic mean free path.

For 2D systems, the mean free path, $l_{in}$, is also determined by the following relation.

$$l_{in} = k_F \cdot \frac{\hbar \mu}{e} \quad (2)$$

which can be derived from the known relations given below.

$$l_{in} = v_F \cdot \tau \quad (3)$$

$$\mu = e\tau/m \quad (4)$$

$$k_F = (2\pi n)^{1/2} \quad (5)$$

$$v_F = \hbar k_F / m \quad (6)$$

where $\mu$ is mobility, $k_F$ is Fermi wave vector, $v_F$ is Fermi velocity, $\tau$ is relaxation time, and $n$ is electron density.

Therefore, using $n$ and $\mu$ found from the Hall data given in TABLE S1 yields a mean free path $l_{in}$ to be 21.6 nm at 10 K, and 7.34 nm at 300 K.

---

* divine.kumah@duke.edu



    

\end{document}